\documentclass[conference]{IEEEtran}
\IEEEoverridecommandlockouts
\usepackage{cite}
\usepackage{amsmath,amssymb,amsfonts}
\usepackage{algorithmic}
\usepackage{graphicx}
\usepackage{textcomp}
\usepackage{xcolor}

\usepackage[hyphens]{url}
\usepackage[hidelinks]{hyperref}
\newcommand{\tabhead}{\textbf}
\usepackage{booktabs,graphicx,makecell,siunitx,eqparbox}

\graphicspath{{./figures/}}
\DeclareGraphicsExtensions{.pdf,.jpeg,.png}

\def\BibTeX{{\rm B\kern-.05em{\sc i\kern-.025em b}\kern-.08em
    T\kern-.1667em\lower.7ex\hbox{E}\kern-.125emX}}
\begin{document}

\title{Comparing Representations for Audio Synthesis Using Generative Adversarial Networks\\
\thanks{This research is supported by the European Union's Horizon 2020  research and innovation program under the Marie Skłodowska-Curie grant agreement No. 765068 (MIP-Frontiers).}
}

\author{
    \IEEEauthorblockN{Javier Nistal}
    \IEEEauthorblockA{
        \textit{Sony Computer Science Laboratories}\\
        Paris, France
    }
\and
    \IEEEauthorblockN{Stefan Lattner}
    \IEEEauthorblockA{
        \textit{Sony Computer Science Laboratories}\\
        Paris, France
    }
\and
    \IEEEauthorblockN{Ga\"{e}l Richard}
    \IEEEauthorblockA{
        \textit{LTCI, T\'{e}lecom Paris} \\
        \textit{Institut Polytechnique de Paris, France}
    }
}

\maketitle

\begin{abstract}
In this paper, we compare different audio signal representations, including the raw audio waveform and a variety of time-frequency representations, for the task of audio synthesis with Generative Adversarial Networks (GANs). We conduct the experiments on a subset of the NSynth dataset. The architecture follows the benchmark Progressive Growing Wasserstein GAN. We perform experiments both in a fully non-conditional manner as well as conditioning the network on the pitch information. We quantitatively evaluate the generated material utilizing standard metrics for assessing generative models, and compare training and sampling times. We show that complex-valued as well as the magnitude and Instantaneous Frequency of the Short-Time Fourier Transform achieve the best results, and yield fast generation and inversion times. The code for feature extraction, training and evaluating the model is available online.\footnote{\url{https://github.com/SonyCSLParis/Comparing-Representations-for-Audio-Synthesis-using-GANs}}
\end{abstract}


\section{Introduction}
In recent years, deep learning for audio has shifted from using hand-crafted features requiring prior knowledge, to features learned from raw audio data or mid-level representations such as the Short-Time Fourier Transform (STFT) \cite{DielemanS14}.
Indeed, this has allowed us to build models requiring less prior knowledge, yet at the expense of data, computational power, and training time \cite{ZhuEH16}.
For example, deep autoregressive techniques working directly on raw audio \cite{OordDZSVGKSK16}, as well as on Mel-scaled spectrograms \cite{melnet}, currently yield state-of-the-art results in terms of quality. However, these models can take up to several weeks to train in a conventional GPU, and also, their generation procedure is too slow for typical production environments. On the other hand, Generative Adversarial Networks (GANs) \cite{Goodfellow2013}, have achieved comparable audio synthesis quality and faster generation time \cite{Engel}, although they still require long training times and large-scale datasets when modeling low or mid-level feature representations \cite{Marafioti, Donahue2018}.

It is still subject to debate what the best audio representations are in machine learning in general, and the best choice may also depend on the respective application and the models employed.
In audio synthesis with GANs, different representations may result in different training and generation times, and may also influence the quality of the resulting output.
For example, operating on representations that compress the information with respect to perceptual principles, or are structured to better support a specific model architecture, may yield faster training and generation times, but may result in worse audio quality.
Therefore, in this paper, we compare different audio signal representations, including the raw audio waveform and a variety of time-frequency representations, for the task of adversarial audio synthesis with GANs. To this end, we adapt several objective metrics, initially developed for the image domain, to audio synthesis evaluation. In addition, we also report on the respective training, generation, and inversion times.
Furthermore, we investigate whether global attribute conditioning may improve the quality and coherence of the generated audio. For that, we perform extensive experimental evaluation when conditioning our models on the pitch information, as well as in a fully unconditional setting. We use a vanilla Progressive Growing Wasserstein GAN built upon convolutional blocks \cite{Karras2017}, as this architecture has achieved state-of-the-art audio synthesis \cite{Engel}.

The paper is organized as follows: In Section 2, we introduce the audio representations used in our experiments. In Section 3, we describe the dataset, architecture design, training procedure, and the metrics used for evaluation. Results are presented in Section 4, and we conclude in Section 5.

\section{Audio representations}
\label{sec:audio_rep}

Audio signals consist of large amounts of data in which relevant information for a specific task is often hidden, and spread over large time spans. Neural Networks can benefit from feeding in sparse representations of the audio data, where few coefficients reveal the information of interest. These types of representations may yield faster training and less complex architectures, which is of particular interest when training deep generative models. Following, we enumerate the audio representations that are compared in this work, highlighting strengths and weaknesses for the specific task of audio synthesis with GANs. Except stated otherwise, we compute the audio representations using Librosa \cite{librosa}.

\begin{itemize}
    \item The \textbf{raw audio waveform} consists of a sequence of numerical samples that specify the amplitude values of the signal at time steps $t$. Using this representation as input is challenging for generative modeling, particularly in the case of music signals \cite{Dieleman2018}. On the other hand, it enables neural networks to build the representation that better suits a specific task without any prior assumptions. In the following, we refer to this representation as \emph{waveform}.

    \item The \textbf{Short-Time Fourier Transform (STFT)} decomposes a signal as a weighted sum of complex sinusoidal basis vectors with linearly spaced center frequencies, unveiling the time-frequency structure of an audio signal. It is commonly decomposed into magnitude and phase components.   
    The latter is typically noisy, which makes it difficult for neural networks to model. This problem is mitigated by using the Instantaneous Frequency (IF), providing a measure of the rate of change of the phase information over time \cite{Boashash}. The STFT transform is cheap to compute and perfectly invertible, which makes it popular for audio synthesis \cite{Engel, Marafioti}. Here we make use of the complex-valued STFT, referred to as \emph{complex} throughout our experiments, as well as the magnitude and IF of the STFT (referred to as \emph{mag-if}).
    
    \item The \textbf{Constant-Q Transform (CQT)} decomposes a signal as a weighted sum of tonal-spaced filters, where each filter is equivalent to a subdivision of an octave \cite{Brown1991}. This musically motivated spacing of frequencies enables representing pitch transpositions as simple shifts along the frequency axis, which is well-aligned with the equivariance property of the convolution operation. The CQT transform has been used as a representation for Music Information Retrieval \cite{Lidy2016CQTBASEDCN} and some works have exploited it for audio synthesis \cite{Philippe}. The main disadvantage of CQT over STFT is the loss of perceptual reconstruction quality due to the frequency scaling in lower frequencies. We use a pseudo invertible CQT \cite{Schrkhuber2010}, as well as an implementation based on the Non-Stationary Gabor Transform (CQ-NSGT)\footnote{\url{https://github.com/grrrr/nsgt}} \cite{Velasco2011}, which allows for perfect reconstruction. In the following, we refer to these two methods for computing the CQT as \emph{cqt} and \emph{cq-nsgt}, respectively.

    \item The \textbf{Mel spectrogram} compresses the STFT in frequency axis by projecting it into a perceptually inspired frequency scale, called the Mel-scale \cite{Stevens1937ASF}. Mel discards the phase information, so we use the iterative method from Griffin and Lim \cite{Griffin1984} to recover the phase for synthesis. We refer to this representation as \emph{mel} throughout our experiments.
 
    \item The \textbf{Mel Frequency Cepstral Coefficients (MFCC)}\cite{Mermelstein80} provide a compact representation of the spectral envelope of an audio signal. Originally developed for speech recognition, they are now widely used in musical applications, as they capture perceptually meaningful musical timbre features \cite{Ravelli}. For synthesis, we invert MFCC to the Mel scale and use Griffin-Lim to recover the phase. We refer to this representation as \emph{mfcc} in our experiments.
\end{itemize}

\section{Experiment setup}\label{sec:typeset_text}

\subsection{Architecture design and training procedure}
\begin{figure}[t]
\centering
\includegraphics[scale=0.5, width=0.8\columnwidth]{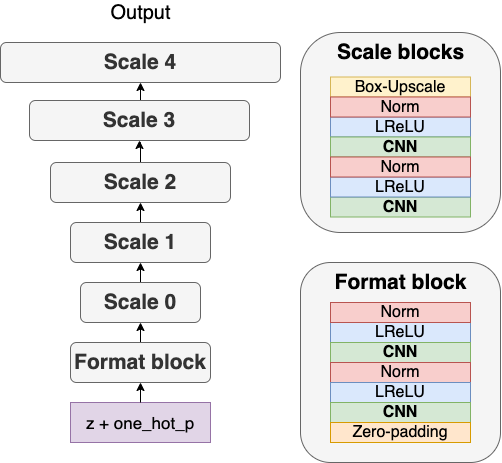}
\caption{The architecture of the generator. The discriminator mirrors this configuration. The format block zero-pads and transforms the input noise and the one-hot attribute encoding to a \((batch\_size, 128, w_0, h_0)\) tensor; where \(w_0\) and \(h_0\) are the sizes of each dimension at the first scale block input. The CNNs within each scale block have 128, 64, 64, 64, 32 feature maps, from low to high resolution, respectively. We apply pixel normalization after every convolutional layer. \label{fig:arch}}
\end{figure}

Our reference architecture is a Progressive Growing GAN (P-GAN) \cite{Karras2017}, borrowed from the Computer Vision literature, which has achieved state-of-the-art results in neural audio synthesis with GANs \cite{Engel}. The generator's architecture is depicted in Figure \ref{fig:arch}. The generator \(G\) samples a random vector \(z\) with 128 components from a  spherical  Gaussian and feeds it together with the one-hot conditional information \emph{one\_hot\_p} through a \emph{Format} block and a stack of \emph{Scale} blocks. The \emph{Format} block  turns the 1D input vector \emph{z + one\_hot\_p}, with size 128 + 27, into a 4D convolutional input by zero-padding in the time and frequency-dimension (i.e., placing the input vector in the middle of the convolutional input with 128 + 27 convolutional maps). The \emph{Scale} blocks are a stack of convolutional and box-up-sampling blocks that transform the convolutional input to the generated output signal. The discriminator \(D\) is composed of convolutional and down-sampling blocks, mirroring the configuration of the generator. \(D\) estimates the Wasserstein distance between the real and generated distributions \cite{Gulrajani2017}. We use a gradient penalty of 10.0 to enforce the Lipschitz constraint and pixel normalization at each layer. We initialize weights to zero and apply He's constant \cite{HeZRS15} for normalizing each layer at run-time in order to ensure an equalized learning rate. Also, we use a mini-batch standard deviation before the last layer of \(D\) \cite{SalimansGZCRCC16}. This encourages \(G\) to generate more variety and thus reduces mode collapse. For conditional model experiments, we add an auxiliary classification loss to the discriminator that learns to predict the pitch label \cite{OdenaOS17}.

Training is divided into phases, wherein each phase a new layer, generating a higher-resolution output, is added to the existing stack. A blending parameter \(\alpha{}\) progressively fades in the gradient derived from the new layers, minimizing possible  perturbation effects. We train all the models for 1.1M iterations on batches of 8 samples: 200k iterations in each of the first four phases and 300k in the last one. We employ Adam as the optimization method.

\subsection{Dataset}
\begin{table}
    \caption{Audio representation configuration}
    \label{tab:audio-rep}
    \begin{center}
        \begin{tabular}{ l c c c}
          \toprule
          \tabhead{Audio rep.} & \tabhead{channels}  & \tabhead{freq. bins} & \tabhead{time frames/instance}  \\
          \midrule
          waveform  & 1  & - & 16000\\
          complex   & 2 & 512 & 64\\
          mag-if    & 2 & 512 & 64\\
          cq-nsgt   & 4 & 97 & 948\\
          cqt       & 2 & 84 & 256 \\
          mel       & 1 & 128 & 64\\
          mfcc      & 1 & 128 & 64\\
          \bottomrule
        \end{tabular}
    \end{center}

\end{table}
\label{subsec:data}
For this work, we make use of the NSynth dataset \cite{Engel2017},  consisting of approximately 300,000  single-note audios played by more than 1,000 different instruments from 10 different families. The samples are aligned, meaning that the onset of each note is centered at time 0. It contains labels for pitch, velocity, instrument type, acoustic qualities (acoustic or electronic), and more, although, for this particular work, we only make use of the pitch information for those experiments regarding conditional models.  Each sample is four seconds long, with a 16kHz sample rate. The subset of NSynth we use here only contains acoustic instruments from the brass, flutes, guitars, keyboards, and mallets families. We also trim down the audio samples from 4 to 1 seconds and only consider samples with a MIDI pitch range from 44 to 70 (103.83 - 466.16 Hz), as this is the range in which there exist the most examples from the chosen instrument types. This yields a subset of approximately 22k sounds with balanced instrument class distribution. For the evaluation, we perform an 80/20\% split of the data.

All time-frequency representations, except \emph{cqt} and \emph{cq-nsgt}, are computed using an FFT size of 1024 and 75\% overlapping. In the case of \emph{mel} and \emph{mfcc}, we employ a filter-bank of 128 Mel bins. For \emph{mfcc}, we do not compress the Mel frequency information so as to preserve pitch information. \emph{cqt} is computed using 12 bins per octave with a total of 84 bins. \emph{cq-nsgt} is computed using 193 bins and assuming a complex signal. This leads to a non-symmetric spectrogram in which correlated frequency information is mirrored around the DC component. In order to make the information more local, we fold the magnitude and phase components and discard the DC. The resulting tensor sizes for each representation are summarized in Table \ref{tab:audio-rep}.


\subsection{Evaluation}
Evaluating generative models is not straight-forward. Particularly in the case of audio synthesis, where the goal of synthesizing perceptually-realistic audio is hard to formalize. A common practice is to compare models by listening to samples and to measure their performance in classification tasks. Similarly to previous work \cite{Engel}, we evaluate our models against a diverse set of metrics that are common in the literature, each capturing a distinct aspect of the model's performance.

\begin{itemize}
    \item The \textbf{Inception Score (IS)} is defined as the mean KL divergence between the conditional class probabilities \(p(y|x)\), and the marginal distribution \(p(y)\) using the predictions of an Inception classifier \cite{SalimansGZCRCC16}:
    \begin{equation}
        \exp{\big (E_x[KL(p(y|x)||p(y))]\big )}
        \label{eq:iscore}
    \end{equation}

    Similar to \cite{Engel}, we adapt this metric to audio evaluation by training the Inception Net\footnote{ \url{https://github.com/pytorch/vision/blob/master/torchvision/models/inception.py}} on the tasks of instrument and pitch classification from magnitude STFT spectrograms. We refer to these as Pitch Inception Score (PIS) and Instrument Inception Score (IIS), respectively. IS penalizes models whose examples are not classified into a single class with high confidence, as well as models whose examples belong to only a few of all the possible classes.
    We trained the pitch and instrument inception model variants on the same sub-set of the NSynth used throughout our experiments, with a train-validation split of 80\% and 20\%, respectively.
    \item \textbf{Kernel Inception Distance (KID)}. The KID measures the dissimilarity between samples drawn independently from real and generated distributions \cite{BinkowskiSAG18}. It is defined as the squared Maximum Mean Discrepancy (MMD) between Inception representations. A lower MMD means that the generated probability distribution \(P_g\) is closer to the real data distribution \(P_r\). We employ the unbiased estimator of the squared MMD \cite{GrettonBRSS12} between \(m\) samples \(X\thicksim P_r\) and \(n\) samples \(Y\thicksim P_g\), for some fixed characteristic kernel function \(k\), defined as:
    
    \begin{footnotesize}
    \begin{equation}
            \begin{split}
                \text{MMD}^2 (X, Y) =
                    & \quad \frac{1}{m(m-1)}\sum_{i\neq j}^{m}k(x_i, x_j) \\ + & \quad \frac{1}{n(n-1)}\sum_{i\neq j}^{n}k(y_i, y_j) \\ - & \quad \frac{2}{mn}\sum_{i=1}^{m} \sum_{j=1}^{n} k(x_i,y_j)
            \end{split}
    \end{equation}
    \end{footnotesize}
    
    Here, we use an inverse multi-quadratic kernel (IMQ) \(k(x,y) = 1 / (1+||x-y||^2 / 2 \gamma^2)\) with \(\gamma^2 = 8\),
    as it has a heavier tail than a Gaussian kernel, hence, it is more sensitive to outliers. We borrow this metric from the Computer Vision literature and apply it to the audio domain.

    \item The \textbf{Fr\'{e}chet Audio Distance (FAD)} compares the statistics of real and fake data computed from an embedding layer of a pre-trained VGGish model\footnote{\url{https://github.com/google-research/google-research/tree/master/frechet\_audio\_distance}} \cite{fad}. Viewing the embedding layer as a continuous multivariate Gaussian, the mean and covariance are estimated for real and fake data, and the FAD between these is calculated as:
    \begin{equation}
        FAD = ||\mu{}_r - \mu{}_g||^2 + tr(\Sigma_r + \mu{}_g - 2\sqrt{\Sigma_r\Sigma_g})
        \label{eq:fad}
    \end{equation}
    where \((\mu{}_r, \Sigma_r)\) and \((\mu{}_g, \Sigma{}_g)\) are the mean and covariances of \(P_r\) and \(P_g\), respectively. Lower FAD means smaller distances between synthetic and real data distributions. FAD performs well in terms of robustness against noise, computational efficiency, consistency with human judgments and sensitivity to intra-class mode dropping.
\end{itemize}
\section{Results}
\subsection{Qualitative results}
We encourage the reader to listen to the audio examples provided at the accompaniment website.\footnote{\url{https://sites.google.com/view/audio-synthesis-with-gans}}
\emph{mag-if} and \emph{complex} seem to have the best-perceived quality, and are comparable to state-of-the-art works on adversarial audio synthesis (e.g., \cite{Engel, Donahue2018}).
We note that every representation has specific artifacts. While \emph{waveform} seems to suffer from general broad-band noise, in $nsgt$ problems in reproducing plausible phase information sometimes lead to percussive artifacts (and frequency sweeps) at the beginning and end of a sample. The samples in other representations suffer from ringing (e.g., \emph{complex}) or from pitch distortion (e.g., \emph{cqt}).

Interpolation between random points in the latent space seems to produce particularly smooth transitions in \emph{complex}, followed by \emph{mag-if}, \emph{cqt}, and \emph{cq-nsgt}.
The model trained on \emph{mel} fails to faithfully reproduce the timbral characteristics of the training data, and also does not generate the required pitches in the pitch-conditional setting (it always produces the same pitch for a given $z$). As the training setup is the same for every representation, the reason for that is not clear.

\subsection{Quantitative results}
\begin{table}
    \caption{Unconditional models. Higher is better for PIS and IIS, lower is better for PKID, IKID and FAD.}
    \label{tab:uncond}
    \begin{center}
        \begin{tabular}{ l l l l l l}
          \toprule
          \tabhead{Models} & \tabhead{PIS}  & \tabhead{IIS} & \tabhead{PKID} & \tabhead{IKID} & \tabhead{FAD}  \\
          \midrule
          real data & 12.5 & 4.0 & 0.000 & 0.000 & 0.01 \\
          waveform  & 3.7 & 1.8 & 0.083 & 0.291 & 6.46 \\
          complex   & \bf{9.5} & 2.8 & \bf{0.007} & 0.124 & 3.17 \\
          mag-if    & 7.3 & 2.7 & 0.015 & 0.149 & 2.71 \\
          cq-nsgt   & 8.1 & \bf{3.4} & 0.012 & \bf{0.041} & \bf{2.11} \\
          cqt       & 7.8 & 2.6 & 0.013 & 0.112 & 2.55 \\
          mel       & 2.3 & 1.1 & 0.147 & 0.300 & 5.20 \\
          mfcc      & 8.9 & 3.0 & 0.008 & 0.080 & 2.92 \\
          \bottomrule
        \end{tabular}
    \end{center}
\end{table}

\begin{table}
    \caption{Conditional models. Higher is better for PIS and IIS, lower is better for PKID, IKID and FAD.}
    \label{tab:cond}
    \begin{center}
        \begin{tabular}{ l l l l l l}
          \toprule
          \tabhead{Models} & \tabhead{PIS}  & \tabhead{IIS} & \tabhead{PKID} & \tabhead{IKID} & \tabhead{FAD}  \\
          \midrule
          real data & 12.5 & 4.0 & 0.000 & 0.000 & 0.01  \\
          waveform  & 3.4  & 2.1 & 0.222 & 0.108 & 1.87 \\
          complex      & 12.0 & 2.7 & 0.005 & 0.159 & \bf{0.11} \\
          mag-if    & \bf{12.6} & \bf{3.9} & \bf{0.002} & \bf{0.020} & 0.12 \\
          cq-nsgt   & 7.6 & 3.3 & 0.014 & 0.049 & 0.12\\
          cqt      & 12.3 & \bf{3.9} & 0.008 & 0.107 & 2.03 \\
          mel       & 12.3 & 3.8 & 0.165 &  0.371 & 4.79 \\
          mfcc     & 9.7 & 3.7 & 0.006 & 0.074 & 2.62 \\
          \bottomrule
        \end{tabular}
    \end{center}
\end{table}

\begin{table}
    \caption{Metrics of post-processed real data for lossy transformations. Higher is better for PIS and IIS, lower is better for PKID, IKID and FAD.}
    \label{tab:bounds}
    \begin{center}
        \begin{tabular}{ l l l l l l}
          \toprule
          \tabhead{Models} & \tabhead{PIS}  & \tabhead{IIS} & \tabhead{PKID} & \tabhead{IKID} & \tabhead{FAD}  \\
          \midrule
          cqt      & 10.5 & 3.1 & 0.001 & 0.001 & 0.66 \\
          mel       & 12.5 & 3.7 & 0.001 &  0.001 & 0.31 \\
          mfcc     & 12.8 & 3.4 & 0.001 & 0.001 & 1.29 \\
          \bottomrule
        \end{tabular}
    \end{center}
\end{table}

The quantitative evaluation for samples generated by the unconditional and conditional models are shown in Table \ref{tab:uncond} and Table \ref{tab:cond}, respectively. We observe a trend that the figures get worse from \emph{complex} and \emph{mag-if} to \emph{mel} and \emph{waveform}. In some metrics, the highest quality models (\emph{complex}, \emph{mag-if}, and \emph{cqt}) obtain results close to the real data. Furthermore, the results are generally better in the conditional setting. This is probably because the pitch-conditioning signal guides the generator in covering the variance over pitches, making it easier for the generator / discriminator pair to learn the remaining variances.
Informal listening tests suggest that PKID, IKID and FAD are better aligned with perceived sound quality than PIS and IIS. In PKID, IKID and FAD (in both, the conditional and unconditional setting), the models of all representations seem to perform similarly, except \emph{mel} and \emph{waveform}, which both yield considerably worse results.

PIS and IIS seem to correspond better with perceived quality in the unconditional setting (with \emph{waveform} and \emph{mel} having low PIS and IIS) than in the conditional setting. In the latter, PIS and IIS fail to reflect the incapability of the model trained on \emph{mel} to produce clear pitches, and to faithfully reproduce the timbral characteristics of the training data. Despite this, we note that both PIS and IIS are high for that model. Conversely, for data generated in the \emph{waveform} domain, the PIS and IIS are low, even though pitch and instrument types can be clearly perceived in informal listening tests. This suggests that the inception models are not robust to the particular artefacts of these representations and therefore not very reliable in measuring the overall generation quality.

For lossy representations (i.e., \emph{cqt}, \emph{mel} and \emph{mfcc}), the quantitative evaluation may suffer from a bias introduced by the lossy compression itself. Therefore, we compute the lower bounds of each representation by encoding/decoding the dataset used for our experiments in the respective transformations, and treating that as ``generated data'' in the evaluation. Table \ref{tab:bounds} shows the results of this experiment. While \emph{cqt} seems to have slightly worse lower bounds in general, the FAD of \emph{mfcc} is worse than that of \emph{mel}, even though there are no audible differences in the audio. Apparently the cosine-transform used to compute \emph{mfcc} from \emph{mel} introduces non-audible artifacts, which have considerable effect on the latent representations of the Inception model.

Table \ref{tab:training_times} shows the training, sampling, and inversion times associated with each model and representation. Note that training times are just rough measures, as they might be affected by variations in performance and resource availability in the training infrastructure.  We can observe that, in general, representations with higher compression yield faster training and sampling times, but at the expense of slower inversion. \emph{cqt} produces the best training, sampling, and inversion times trade-off, followed by the \emph{complex} and \emph{mag-if} representations.

\begin{table}
    \caption{Training, sampling and inversion times for each model}
    \label{tab:training_times}
    \begin{center}
        \begin{tabular}{ l c c c}
          \toprule
          \tabhead{Models} & \tabhead{training (days)}  & \tabhead{sampling (s)} & 
          \tabhead{inversion (s)} \\
          \midrule
          waveform  & 6.1 & 1.31 & 0.00\\
          complex      & 3.5 & 0.20 & 0.01\\
          mag-if    & 4.5 & 0.24 & 0.02\\
          cq-nsgt   & 5.3 & 0.46 & 0.03 \\
          cqt      & 2.1 & 0.09 & 0.03 \\
          mel       & 1.5 & 0.04 & 3.69 \\
          mfcc      & 2.0 & 0.07 & 10.80 \\
          \bottomrule
        \end{tabular}
    \end{center}
    \vspace{-0.3cm}
\end{table}

\section{Conclusion}
In this work, we compared a variety of audio representations for the task of adversarial audio synthesis of pitched sounds. We performed quantitative and qualitative evaluation, and reported on training, generation, and inversion times. We found that \emph{complex} and \emph{mag-if} yield the best quantitative metrics, which is also aligned with informal listening of the generated samples. This is interesting, as we are not aware that \emph{complex} was used before in audio generation. We also found that evaluation metrics are generally aligned with perceived quality, but in some cases they can be sensitive to non-audible representation-specific artifacts (e.g., FAD), or yield figures which seem over-optimistic when listening to the examples (e.g., PIS and IIS).






\bibliographystyle{IEEEtran}  
\bibliography{bibliography}


\end{document}